\documentclass[12pt,english]{article}
\usepackage[T1]{fontenc}
\usepackage[latin9]{inputenc}
\usepackage{geometry}
\usepackage{babel}
\usepackage{float}
\usepackage{amsmath}
\usepackage{graphicx}
\usepackage{setspace}
\usepackage[authoryear]{natbib}
\usepackage[unicode=true,pdfusetitle,
 bookmarks=true,bookmarksnumbered=false,bookmarksopen=false,
 breaklinks=false,pdfborder={0 0 0},pdfborderstyle={},backref=false,colorlinks=false]
 {hyperref}

\makeatletter

\providecommand{\tabularnewline}{\\}

\usepackage{setspace}
\usepackage{lineno}
\usepackage{natbib}
\usepackage{pdflscape}
\usepackage{hyperref}
\usepackage{multirow}
\usepackage{amssymb}
\usepackage{xcolor}

\g@addto@macro\normalsize{%
  \setlength\abovedisplayskip{1pt}
  \setlength\belowdisplayskip{4pt}
  \setlength\abovedisplayshortskip{2pt}
  \setlength\belowdisplayshortskip{2pt}
}
\usepackage{titlesec}
\titlespacing*{\section}{0pt}{3ex plus 2ex}{1ex}
\titleformat*{\section}{\fontsize{14}{14}\bfseries}

\titlespacing*{\subsection}{0pt}{3ex plus 2ex}{1ex}
\titleformat*{\subsection}{\fontsize{12}{12}\bfseries}

\makeatother

\begin{document}
\begin{flushleft}
\setcounter{footnote}{0}
\begin{flushleft}Running Head: \textbf{Tree shrinkage priors}
\par\end{flushleft}

\noindent \textbf{Model selection for ecological community data using
tree shrinkage priors}

\begin{doublespace}
\medskip{}

\end{doublespace}

\begin{singlespace}
\textbf{Trevor J. Hefley}

{\small{}Department of Statistics}{\small\par}

{\small{}Kansas State University}{\small\par}
\end{singlespace}

\medskip{}

\begin{singlespace}
\textbf{Haoyu Zhang}

{\small{}Department of Statistics}{\small\par}

{\small{}Kansas State University}{\small\par}
\end{singlespace}

\begin{doublespace}
\medskip{}

\end{doublespace}

\begin{singlespace}
\textbf{Brian R. Gray}

U.S. Geological Survey

Upper Midwest Environmental Sciences Center

La Crosse, Wisconsin

brgray@usgs.gov
\end{singlespace}

\begin{doublespace}
\medskip{}

\end{doublespace}

\begin{singlespace}
\textbf{Kristen L. Bouska}

U.S. Geological Survey

Upper Midwest Environmental Sciences Center

La Crosse, Wisconsin

kbouska@usgs.gov

\bigskip{}

\textbf{Contribution type:} Article

\textbf{Word count:} 304 (abstract); 10,763 (total)

\textbf{Number of references: }72

\textbf{Number of figures: }5

\textbf{Number of tables: }1\medskip{}

\pagebreak{}

\bigskip{}

\textbf{Corresponding author: }

Trevor J. Hefley

{\small{}Department of Statistics}{\small\par}

205 Dickens Hall 

1116 Mid-Campus Drive North 

Manhattan, KS 66506

(785)-532-0703

thefley@ksu.edu
\end{singlespace}

\bigskip{}

\begin{singlespace}
\textbf{Statement of authorship: }T.J.H, B.R.G and K.L.B conceived
the study. T.J.H and H.Z developed the statistical methods and conducted
the data analysis. T.J.H and H.Z wrote the manuscript. All authors
contributed substantially to revisions. \bigskip{}

\textbf{Data accessibility statement:} The aquatic vegetation and
fisheries data are publicly available from \citet{UMRR}, but the
specific subsets used in this study will be archived in the Dryad
Digital Repository. For the purpose of peer review, the data have
been submitted as supporting material as a compressed file named Data.zip.

\bigskip{}

\textbf{Reproducibility statement:} Annotated computer code capable
of reproducing all results and figures associated with the aquatic
vegetation and fisheries data examples are provided in Appendix S1
and S2.\bigskip{}

\end{singlespace}

\setlength{\parindent}{0.7cm}

\pagebreak{}

\section*{Abstract}

Researchers and managers model ecological communities to infer the
biotic and abiotic variables that shape species' ranges, habitat use,
and co-occurrence which, in turn, are used to support management decisions
and test ecological theories. Recently, species distribution models
were developed for and applied to data from ecological communities.
Model development and selection for ecological community data is difficult
because a high level of complexity is desired and achieved by including
numerous parameters, which can degrade predictive accuracy and be
challenging to interpret and communicate. Like other statistical models,
multi-species distribution models can be overparameterized. Regularization
is a model selection technique that optimizes predictive accuracy
by shrinking or eliminating model parameters. For Bayesian models,
the prior distribution automatically regularizes parameters. We propose
a tree shrinkage prior for Bayesian multi-species distributions models
that performs regularization and reduces the number of regression
coefficients associated with predictor variables. Using this prior,
the number of regression coefficients in multi-species distributions
models is reduced by estimation of unique regression coefficients
for a smaller number of guilds rather than a larger number of species.
We demonstrated our tree shrinkage prior using examples of presence-absence
data for six species of aquatic vegetation and relative abundance
data for 15 species of fish. Our results show that the tree shrinkage
prior can increase the predictive accuracy of multi-species distribution
models and enable researchers to infer the number and species composition
of guilds from ecological community data. The desire to incorporate
more detail and parameters into models must be tempered by the limitations
of the data and objectives of a study, which may include developing
models that have good predictive accuracy and output that is easier
to share with policymakers. Our tree shrinkage prior reduces model
complexity while incorporating ecological theory, expanding inference,
and can increase the predictive accuracy and interpretability of Bayesian
multi-species distribution models. \bigskip{}

\begin{doublespace}
\noindent \textbf{\textit{\emph{Key-words: }}}Bayesian statistics,
community ecology, ecological guild, joint species distribution model,
model averaging, model selection, multi-species distribution model,
prior distribution, regularization\vspace{-1cm}

\end{doublespace}

\section*{Introduction}

Over the past 20 years species distribution models (SDMs) have become
one of the most widely used quantitative methods in ecology (\citealt{elith2009species,franklin2010mapping,guisan2017habitat}).
Species distribution models are routinely fit to presence-only, presence-absence,
and abundance data to understand the biotic and abiotic variables
that influence the habitat use and geographic distribution of a species
(\citealt{aarts2012comparative,mcdonald2013location,hefleyHSDM}).
The output of SDMs are used in scientific studies to test ecological
theories and in application to delineate areas of high conservation
value which informs management decisions (e.g., Wisz et al. \citeyear{wisz2013role};
\citealt{guisan2013predicting,hefley2015use}). Regardless of the
type of data or method, the ultimate goal when using SDMs includes
making reliable inference and accurate predictions.

In many applications, a SDM is fit to data from a single species.
If the study objectives require the analysis of data from multiple
species, then a SDM is fit to data from each species and the models
are combined using stacking (Norberg et al. \textit{in press} \nocite{ecomonograph2019};
\citeauthor{zurelltesting} \textit{in press}). Although this species-by-species
approach is common it presents several challenges that limit the ability
of researchers and managers to make reliable inference and accurate
predictions (\citealt{hui2013mix,madon2013community,taylor2017joint}).
For example, if the goal of a study is to produce accurate range maps
then spatial predictions for a species that is rare will lack precision.
Imprecise spatial predictions result from parameter estimates that
have a high variance, which occurs because parameters are estimated
using limited data from only a single rare species (e.g., \citealt{hefley2015existence}).
As another example, consider the case where the goal is to infer the
influence of a predictor variable such as temperature. Fitting a SDM
for each species requires estimating a unique temperature coefficients
for each species. Even if there are only a few predictor variables
it will be difficult to summarize the regression coefficients for
a moderate number of species, which will hinder communication with
policymakers.

Recently, SDMs for multiple species have been developed. So-called
joint- or multi-species distribution models (MSDMs) enable a single
model to be fit simultaneously to data from multiple species; the
benefits include: 1) community-level inference and resulting prediction;
2) incorporation of biotic interactions; 3) more accurate predictions
of single-species' distributions due to ``borrowing of strength''
across data from co-occurring species; and 4) the potential for model
simplification.

Many types of MSDMs have been developed using statistical or machine
learning approaches (see synthesis in Norberg et al. \textit{in press}
\nocite{ecomonograph2019} and \citealt{wilkinson2019comparison}).
Hierarchical Bayesian modeling is a statistical approach that is easily
customizable and widely used to model the distribution of multiple
species (e.g., \citealt{taylor2017joint,johnson2017modeling,lany2018asymmetric,schliep2018joint,ovaskainen2017make,wilkinson2019comparison}).
Briefly, hierarchical Bayesian models are specified using conditional
probability distributions that represent the data collection process
(i.e., the ``data model''), the latent ecological process (i.e.,
the ``process model''), and prior knowledge about parameters (i.e.,
the ``parameter model''; \citealt{wikle2019spatio}, pgs. 11\textendash 13;
\citealt{hobbs2015bayesian}, ch. 6). 

For presence-absence data, \citet{wilkinson2019comparison} describes
the quintessential hierarchical Bayesian MSDM. This includes the data
model
\begin{equation}
y_{i,j}|z_{i,j}\sim\text{Bern(}g^{-1}(z_{i,j}))\,,
\end{equation}
where $y_{i,j}=1$ if the $i^{\text{th}}$ site ($i=1,2,...,n$) is
occupied by the $j^{\text{th}}$ species ($j=1,2,...,J$) and $y_{i,j}=0$
if absent, ``Bern'' is a Bernoulli distribution, $g^{-1}(\cdot)$
is an inverse link function (e.g., logistic, probit), and $z_{i,j}$
is the latent process on the link scale. For MSDMs, a widely used
process model is
\begin{equation}
\mathbf{z}|\boldsymbol{\mu},\ensuremath{\boldsymbol{\Sigma}}\sim\text{N(\ensuremath{\boldsymbol{\mu}}\ensuremath{,}\ensuremath{\boldsymbol{\Sigma}})\,,}
\end{equation}
where $\mathbf{z}$ is a random vector with elements $z_{i,j}$, $\boldsymbol{\mu}$
is expected value (mean), and $\ensuremath{\boldsymbol{\Sigma}}$
is the variance-covariance matrix of a multivariate normal distribution.
The vector, $\boldsymbol{\mu}$, has elements $\mu_{i,j}$ and is
typically specified using
\begin{equation}
\mu_{i,j}=\alpha_{j}+\mathbf{x}_{i}^{'}\boldsymbol{\beta}_{j}\,,
\end{equation}
where, for the $j^{\text{th}}$ species, $\alpha_{j}$ is the intercept
and $\boldsymbol{\beta}_{j}$ are the $K$ regression coefficients
(i.e., $\boldsymbol{\beta}_{j}\equiv(\beta_{j,1},\beta_{j,2},...,\beta_{j,K})^{'}$).
The vector $\mathbf{x}_{i}$ contains $K$ measured predictor variables
at the $i^{\text{th}}$ site, which are the same for all $J$ species
(i.e., $\mathbf{x}_{i}\equiv(x_{i,1},x_{i,2},...,x_{i,K})^{'}$).
In Eq. 2, the process model is specified jointly (i.e., using a multivariate
distribution) to enable sharing of information across species.

In previous studies, the variance-covariance matrix, $\boldsymbol{\Sigma}$
in Eq. 2, accounted for residual autocorrelation due to space, time,
and/or biotic interactions. For example, within the literature on
MSDMs, a main focus of methodological developments has been on parametrizing
$\boldsymbol{\Sigma}$ to account for biotic factors, such as species
co-occurrence, that can not be explained by the measured predictor
variables $\mathbf{x}_{i}$ (e.g., \citealt{warton2008penalized,warton2015so,ovaskainen2017make,niku2017generalized,wilkinson2019comparison}).
By comparison, however, there are only a few studies that have focused
on the parameterization of $\boldsymbol{\mu}$ in Eq. 2 (e.g., \citealt{johnson2017modeling}).

Regardless of the type of data, most MSDMs use a linear equation to
model the influence of predictor variables. For example, all seven
methods compared by \citet{wilkinson2019comparison} use Eq. 3 which,
if there are $J$ species and $K$ predictor variables, will result
in $J\times K$ regression coefficients. For even a small number of
predictor variables and species, the overabundance of regression coefficients
will be challenging to summarize and interpret and may result in a
MSDM that is overparameterized.

For many Bayesian MSDMs, the parameter model or prior is specified
using simple distributions. For example, in six of the seven methods
presented by \citet{wilkinson2019comparison} priors for the regression
coefficients were specified using uniform distributions or independent
normal distributions with known hyperparameters (e.g., $\beta_{j,k}\sim\text{N}(0,10)$).
Although using simple parameter models is common practice, more sophisticated
parameter models are useful. For example, the seventh method in \citet{wilkinson2019comparison}
uses a multivariate normal distribution with unknown hyperparameters
as a parameter model which enables estimation and inference regarding
the correlation among regression coefficients. For MSDMs, the notion
of a model for the parameters is important because it can enable novel
inference and perform regularization. Regularization is a model selection
technique used to optimize predictive accuracy by controlling model
complexity (\citealt{bickel2006regularization,hooten2015guide}). 

\citet{johnson2017modeling} develop a hierarchical Bayesian MSDMs,
which includes a parameter model that leverages the concept of ecological
guilds (\citealt{simberloff1991guild}). For MSDMs, the ecological
guild concept is useful because it provides a framework to incorporate
ecological theory and reduce the number of regression coefficients
associated with predictor variables. For example, the ecological guild
concept suggests that some species may respond to predictor variables
in a similar direction and magnitude as a result of similar resource
use or  ecological role. In other words, species within the same guild
may be associated with similar values of the regression coefficients
in Eq. 3. If the regression coefficients for two or more species are
effectively the same, then the complexity of the MSDM can be reduced
by estimating unique regression coefficients for a smaller number
of guilds rather than a larger number of species. 

If the guild membership is known, then incorporating the guild concept
into MSDMs is trivial and involves using Eq. 3 with a guild by predictor
variable interaction effect rather than a species by predictor variable
(\citealt{johnson2017modeling}). In nearly all applications, the
number and species compositions of the guilds is unknown and must
be estimated from data which can be accomplished using model-based
techniques. For example, as a heuristic imagine that the species-specific
regression coefficients for a single predictor variable in Eq. 3 were
known. In this example, the number and species composition of guilds
could be determined by classifying or clustering the regression coefficients
into homogeneous groups. The classification model would be easier
to interpret when compared to the larger number of species-specific
regression coefficients because it provides a data-driven approach
to determine which species can be modeled with a single guild-level
regression coefficient. This ad-hoc approach is a useful heuristic
because it gives insight into how a large number of regression coefficients
can be summarized using a model for the parameters. Within a hierarchical
Bayesian framework, formalizing this concept is straightforward by
specifying an appropriate parameter model. 

Simple independent parameter models that are commonly used for MSDMs,
such as $\beta_{j,k}\sim\text{N}(0,10)$, do not leverage the information
in the data shared among species. The independence assumption eliminates
the possibility that information about parameters will be shared across
species. Shared information among the species, however, can be accessed
by incorporating the guild concept into the parameter model. For example,
\citet{johnson2017modeling} developed a hierarchical Bayesian MSDM
using a Dirichlet process mixture for the parameter model, which enables
regression coefficients to be estimated using data from all species
within the same guild. Briefly, the Dirichlet process mixture is a
distribution that can be used as a parameter model to cluster the
species into guilds while simultaneously fitting a MSDM to community
data. 

Classification and regression trees (CART; \citealt{CART1984}) are
a machine learning approach with widespread use in ecology (\citealt{de2000classification,de2002multivariate,elith2008working}).
For ecological data, CART offer a semi-automated approach to build
predictive models that are easy to interpret. Although Bayesian variants
of CART have been available for some time (e.g., \citealt{chipman1998bayesian,denison1998bayesian}),
they are rarely used by ecologists. This is perhaps because accessible
versions of Bayesian CART software are not easily modified to accommodate
ecological data. For example, current implementations of Bayesian
CART do not include a ``data model,'' which is needed to account
for imperfect measurements of ecological processes. Conceptually,
CART can be embedded within hierarchical Bayesian models (at any level),
which would enable ecologists to exploit the benefit of both approaches
(\citealt{shaby2012embedding}).

In our work, we show how a specific type of CART can be used as a
parameter model for Bayesian MSDMs. Similar to \citet{johnson2017modeling},
our approach leverages the concept of ecological guilds but reduces
model complexity via a novel tree shrinkage prior (TSP). We illustrate
the TSP by fitting MSDMs to presence-absence and relative abundance
data collected to inform management of freshwater aquatic vegetation
and fish populations. These data are collected as part of a long-term
environmental monitoring program intended to inform management of
the Upper Mississippi River System. Managers are required to use these
data to make and justify management recommendations. This requires
understanding the predictor variables that influence the distribution
of a large number of species using models that can be easily shared
with policymakers but that are capable of making accurate predictions.
\begin{spacing}{1.9}

\section*{Materials and methods}
\end{spacing}
\begin{spacing}{1.9}

\subsection*{INCORPORATING GUILDS INTO MULTI-SPECIES DISTRIBUTION MODELS}
\end{spacing}

\noindent The mechanics of incorporating guilds into hierarchical
Bayesian MSDMs was presented by \citet{johnson2017modeling}. For
MSDMs, this involves a simple modification of Eq. 3 
\begin{equation}
\mathbf{\boldsymbol{\mu}}_{i}=\mathbf{\boldsymbol{\alpha}}+(\mathbf{x}_{i}^{'}\otimes\mathbf{Z})\boldsymbol{\gamma}\,,
\end{equation}
where, $\mathbf{\boldsymbol{\mu}}_{i}\equiv(\mu_{i,1},\text{\ensuremath{\mu_{i,2}},...,\ensuremath{\mu_{i,J}}})^{'}$
is a vector that specifies the expected value for the process model
at the $i^{\text{th}}$ site for $J$ species, $\boldsymbol{\alpha}\equiv(\alpha_{1},\alpha_{2},...,\alpha_{J})^{'}$
contains intercept parameters for each species, and $\mathbf{x}_{i}\equiv(x_{i,1},x_{i,2},...,x_{i,K})^{'}$
contains $K$ predictor variables measured at the $i^{\text{th}}$
site. The matrix $\mathbf{Z}$ has dimensions $J\times G$ where the
$j^{\text{th}}$ row contains an indicator variable that links the
predictor variables to the guild-specific regression coefficients
contained within the vector $\boldsymbol{\gamma}$. The mathematical
symbol $\otimes$ denotes a Kronecker product. For example, let there
be $J=4$ species, $G=2$ guilds, and $K=2$ predictor variables,
then for the $i^{\text{th}}$ site, one configuration of Eq. 4 is
\begin{equation}
\begin{bmatrix}\mu_{i1}\\
\mu_{i2}\\
\mu_{i3}\\
\mu_{i4}
\end{bmatrix}=\begin{bmatrix}\alpha_{1}\\
\alpha_{2}\\
\alpha_{3}\\
\alpha_{4}
\end{bmatrix}+\left(\begin{bmatrix}x_{i1} & x_{i2}\end{bmatrix}\otimes\begin{bmatrix}1 & 0\\
1 & 0\\
0 & 1\\
0 & 1
\end{bmatrix}\right)\begin{bmatrix}\gamma_{1,1}\\
\gamma_{2,1}\\
\gamma_{1,2}\\
\gamma_{2,2}
\end{bmatrix}\,,
\end{equation}
where the first and second species are members of the first guild
and the third and fourth species are members of the second guild.
The guild-specific regression coefficients are $\boldsymbol{\gamma}\equiv(\gamma_{1,1},\gamma{}_{2,1},\gamma_{1,2},\gamma_{2,2})^{'}$
where the subscripts, $\gamma_{g,k}$, indicate the $g^{\text{th}}$
guild and the $k^{\text{th}}$ predictor variable. 

Modifications to the example given in Eq. 5, illustrates that incorporating
the guild concept into MSDM results in a model that is flexible and
includes two special cases. For example, if $\mathbf{Z}$ is a $4\times1$
matrix with all elements equal to one (i.e., $\mathbf{Z}\equiv(1,1,1,1)^{'}$),
then all species would be in a single guild. Similarly if $\mathbf{Z}$
is a $4\times4$ identity matrix (i.e., a matrix with diagonal elements
equal to one and zero on the off diagonal elements), then each species
would be assigned to a unique guild; this results in the commonly
used MSDM with unique species-specific regression coefficients (i.e.,
Eq. 3) and demonstrates that incorporating the guild concept into
MSDMs can only improve the predictive accuracy.

If the number and species composition of the guilds was known, then
the matrix $\mathbf{Z}$ would also be known. In practice, however,
the number and species composition of the guilds are unknown, which
requires estimation of $\mathbf{Z}$. In turn, $\mathbf{Z}$ regularizes
the MSDM because $\mathbf{Z}$ shrinks and reduces the number of regression
coefficients from $J\times K$ to $G\times K$.
\begin{spacing}{1.9}

\subsection*{\vspace{0.66cm}
TREE SHRINKAGE PRIOR}
\end{spacing}

\noindent A general introduction to tree-based methods is given in
\citeauthor{james2013introduction} (2013; ch. 8) while \citet{linero2017review}
provides a technical review of Bayesian tree approaches. We use vocabulary
found in both \citeauthor{james2013introduction} (2013; ch. 8) and
\citet{linero2017review}. The idea of a TSP is not new but, to our
knowledge, has been used in only one unrelated setting (\citeauthor{guhaniyogi2018large}
\textit{in press}). For MSDMs, the idea is simple; the species within
terminal nodes of the tree represent the guilds and determines the
matrix $\mathbf{Z}$ in Eq. 4. Each terminal node is associated with
guild-level regression coefficients for the predictor variables. For
example, in what follows, we use presence-absence data for six species
of aquatic vegetation and a single predictor variable (water depth).
An example of a binary tree is given in Fig. 1 and shows which species
share the same numerical value of the regression coefficients for
the predictor variable water depth. In this example, the standard
MSDM results in six regression coefficients (i.e., one for each species).
For the predictor variable water depth, a MSDM specified using the
tree shown in Fig. 1 results in only three regression coefficients.
Recall that there are intercept parameters for each species, as a
result the binary tree in Fig. 1 reduces the number of parameters
in the MSDM from 12 to 9. 

There are a variety of techniques to construct tree priors discussed
within the statistics and machine learning literature (\citealt{linero2017review}).
Most techniques rely on a tree generating stochastic process (\citealt{linero2017review}).
Binary trees, like the example in Fig. 1, are a type of stochastic
branching process (\citealt{dobrow2016introduction}, ch. 4). Specification
of a tree generating stochastic process involves specifying probability
distributions that generate a tree structure as a random variable.
For example, a simple stochastic binary tree generating process for
the six species of aquatic vegetation in Fig. 1 involves specifying
a distribution that determines the splitting of parent nodes and a
second distribution that determines the allocation of species to the
child nodes (e.g., in Fig. 1 super guild 1 is a parent node with guilds
1 and 2 as child nodes). For the aquatic vegetation example, one way
to specify a binary tree generating stochastic process would involve
using a ``splitting distribution'' that determines if a parent node
should be split into two child nodes. If a draw from the splitting
distribution results in a value of one, the parent node is split and
draws from a ``assignment distribution'' allocates the species within
the parent node to one of the two child nodes. In practice, the assignment
distribution is $\text{Bern}(0.5)$ whereas the splitting distribution
is $\text{Bern}(p_{split})$, where the hyperparameter $p_{split}$
controls the model complexity by inducing a prior on the guilds. For
example, if $p_{split}=1$ then a binary tree is generated with six
terminal nodes (guilds) that contain only a single species; when $p_{split}=0$
a tree with no splits is generated and all six species are in the
same guild. Finally, the terminal nodes of the MSDM with a TSP has
guild-level regression coefficients, $\gamma_{g,k}$, which require
the specification of priors. For tree-based methods, simple models
like $\gamma_{g,k}\sim\text{N(0,10)}$ are commonly used, (\citealt{linero2017review}),
however, more complex models have been developed (e.g., \citealt{gramacy2008bayesian}).

In sum, specifying a stochastic tree generating process results in
an implied prior on the tree structure. For example, a so-called non-informative
prior could be constructed by specifying a tree generating stochastic
process that generates all possible combinations of guilds with equal
probability (e.g., \citealt{atkinson1992generating}). For the aquatic
vegetation example such a prior would result in 63 unique guilds.
For this example, the MSDM with the TSP could be implemented using
Bayesian model averaging (\citealt{oliver1995pruning}). Model averaging
is a well-known technique among ecologists (e.g., \citealt{hobbs2015bayesian};
Dormann et al. 2018\nocite{dormann2018model}), however, this would
require fitting 63 models (i.e., a model for each unique guild). Although
we could fit a MSDM with the TSP to the aquatic vegetation data using
Bayesian model averaging, this approach would not work for a larger
number of species. For example, the fisheries data used to illustrate
the TSP contains 15 species which results in 32,767 unique guilds.
Despite advances in computational statistics, implementation of tree-based
methods using Bayesian model averaging is challenging because of the
potential for a large number trees (\citealt{chipman2001practical,hooten2015guide,BB2L},
ch. 15). In the next section, we describe alternative strategies for
model fitting that may be less familiar among ecologists.

\subsection*{\vspace{0.66cm}
MODEL FITTING}

Hierarchical Bayesian MSDMs can be fit to presence-absence, abundance,
and presence-only data using a variety of algorithms. Markov chain
Monte Carlo (MCMC) algorithms are a type of stochastic sampling approach
routinely used by ecologists and implemented in standard software
such as WinBugs, JAGS, and Stan (\citealt{BB2L}). As such, we demonstrate
the TSP using MCMC.

Compared with other commonly used ecological models, MCMC based implementation
of Bayesian trees usually requires highly specialized algorithms.
As a result, an inability to easily construct efficient algorithms
has likely slowed the adoption of Bayesian regression trees (\citealt{linero2017review}).
For example, Bayesian model averaging, used to fit Bayesian trees
to data, requires specialized algorithms (e.g., \citealt{oliver1995pruning,hernandez2018bayesian}).
As mentioned, however, this may be computationally challenging or
infeasible for the TSP when there is a large number of guilds. To
increase computational efficiency, search algorithms have been used
to identify higher probability tree structures, however, as noted
by \citet{chipman1998bayesian}, the ``procedure is a sophisticated
heuristic for finding good models rather than a fully Bayesian analysis,
which presently seems to be computationally infeasible.'' Significant
progress continues in the development of computationally efficient
implementations of Bayesian trees (\citealt{linero2017review}), however,
fitting tree-based models to data requires matching the model to computational
algorithms which can be challenging for non-experts.

An appealing implementation for ecologists is presenting by \citet{shaby2012embedding}.
The \citet{shaby2012embedding} approach enables off-the-shelf software
for tree-based methods, such as those available in R, to be embedded
within hierarchical Bayesian models. This is appealing to ecologists
because it makes the TSP accessible by eliminating the need to develop
custom software to implement the TSP. Briefly, the technique of \citet{shaby2012embedding}
is an approximate Gibbs sampler that enables machine learning algorithms
to be embedded within hierarchical Bayesian models. In what follows
we use a model-based recursive partitioning approach that is available
in the R package ``partykit'' and implemented within the \texttt{lmtree(...)}
function (\citealt{hothorn2006unbiased,zeileis2008model}; \citeauthor{zeileisparties}
2019). Model-based recursive partitioning estimates a binary tree
and consequently $\mathbf{Z}$ and $\boldsymbol{\gamma}$ in Eq. 4.
The approximate Gibbs sampler of \citet{shaby2012embedding} is similar
to an empirical Bayesian step within a Gibbs sampler because priors
for a portion of the unknown model parameters are replaced with estimates
(\citealt{casella2001empirical}). As a result, when using the \citet{shaby2012embedding}
approach, there are no distributions or hyperparameters associated
with the TSP that must be specified. 

In what follows, we use standard MCMC algorithms for MSDMs, but employ
the \citet{shaby2012embedding} approach. Constructing the approximate
Gibbs sampler using the \citet{shaby2012embedding} approach for the
aquatic vegetation data example is incredibly simple and requires
only six lines of R code and the ability to sample univariate normal
and truncated normal distributions (e.g., using the R functions \texttt{rnorm(...)}
and \texttt{rtruncnorm(...)}; see section 2.3 in Appendix S1). To
assist readers implementing similar MSDMs with the TSP, we provide
tutorials with the computational details, annotated computer code,
and the necessary code to reproduce all results and figures related
to the analysis for the aquatic vegetation data and fisheries data
examples (see Appendix S1 and S2). \vspace{0.66cm}

\subsection*{STATISTICAL INFERENCE}

A benefit of Bayesian inference is that uncertainty regarding the
structure of the binary tree, and therefore the number and species
composition of the guilds, is automatically accounted for. Under the
Bayesian paradigm, samples from the posterior distribution of the
trees are obtained during the model fitting process. The posterior
distribution of the trees can be summarized to infer characteristics
of the guilds. For example, the species composition of the guilds
with the highest probability (given the data) can be obtained from
the posterior distribution of the trees.

In addition, posterior distributions of the species-level and guild-level
regression coefficients can be sampled and summarized. The species-level
regression coefficients, akin to $\boldsymbol{\beta}$ in Eq. 3, are
a derived quantity calculated using
\begin{equation}
\boldsymbol{\beta}_{j}^{(m)}=\mathbf{W}_{j}^{(m)}\boldsymbol{\gamma}^{(m)}\,,
\end{equation}
where the superscript $m$ is the current MCMC sample, $\boldsymbol{\beta}_{j}$
is vector of regression coefficients for the $j^{\text{th }}$ species,
$\mathbf{W}_{j}$ is a matrix that contains indicator variables linking
the guild-level coefficients to the species. In the presence of guild
membership uncertainty, the species-level regression coefficients
result in a posterior distribution that is a mixture of the guild-level
regression coefficients.

When using the TSP, the posterior distribution of the regression coefficients
for a species are akin to the model averaged posterior distribution
obtained from fitting and averaging multiple MSDMs with differing
species composition and numbers of guilds to the data. Similarly the
guild-level regression coefficients are akin to the regression coefficients
obtained from a single MSDM using a fixed guild structure. We recommend
using the species-level regression coefficients to make inference
because, with a finite amount of data, the species composition of
the guilds will always be uncertain. Inference using the guild-level
regression coefficients is conditional (i.e., valid for a fixed number
and species composition of the guilds) and does not fully account
for uncertainty.

\subsection*{\vspace{0.66cm}
LONG-TERM RESOURCE MONITORING DATA}

\noindent The Mississippi River is the second longest river in North
America and flows a distance of 3,734 km. The Upper Mississippi River
System includes approximately 2,100 km of the Mississippi River channel
and drains a basin covering 490,000 $\text{km}^{2}$ that contains
30 million people. The Upper Mississippi River Restoration Program
was authorized under the Water Resources Development Act of 1986 (\citet{waterlaw}).
The U.S. Army Corps of Engineers provides guidance, has overall Program
responsibility, and established a long term resource monitoring (LTRM)
element that includes collecting biological and geophysical data.
The LTRM element is implemented by the U.S. Geological Survey, Upper
Midwest Environment Sciences Center, in cooperation with the five
Upper Mississippi River System states of Illinois, Iowa, Minnesota,
Missouri, and Wisconsin (\citealt{UMRR}). The expressed goal of the
LTRM element is to ``support decision makers with the information
and understanding needed to maintain the Upper Mississippi River System
as a viable multiple-use large river ecosystem.'' The biological
data collected by LTRM contains a large number of species and exemplifies
the need for interpretable MSDMs with results that are easy to communicate
to policymakers. For example, data from 154 species, some of which
are rare or endangered, have been collected by the LTRM element since
1993; inferences from these data provide context for management recommendations
for these or collections of these species.

As part of LTRM element, data on aquatic vegetation and fish are collected
at multiple sites within six study reaches of the Upper Mississippi
River. To illustrate our method, we use aquatic vegetation and fisheries
data from navigation pool 8 collected in 2015. Complete details of
data collection procedures for the aquatic vegetation data and fisheries
data are given by \citet{vegsampling} and \citet{fishsampling} respectively.
\begin{spacing}{1.9}

\subsection*{\vspace{0.66cm}
EXAMPLE 1: PRESENCE-ABSENCE OF AQUATIC VEGETATION}
\end{spacing}

\noindent Our objective for the first example is to illustrate the
TSP using a simple MSDM. As such, this example was designed to demonstrate
the proposed methods rather than to provide scientifically valid inference.
As such, we use only a single predictor variable (water depth) and
presence-absence absence data of six species of aquatic vegetation.
At 450 unique sites, the presence or absence of a species was recorded.
For our example, we use data for broadleaf arrowhead (\textit{Sagittaria
latifolia}), Canadian waterweed (\textit{Elodea canadensis}), coontail
(\textit{Ceratophyllum demersum}), stiff arrowhead (\textit{Sagittaria
rigida}), white water lily (\textit{Nymphaea odorata}), and wild celery
(\textit{Vallisneria americana}). This resulted in a total of 2,700
observations. For our analysis, we randomly selected 225 sites and
used data from these for model fitting. We used data from the remaining
225 sites to test the predictive accuracy of the models. 

For the aquatic vegetation data, we use the data model
\begin{equation}
\mathbf{y}_{i}|\ensuremath{\mathbf{\boldsymbol{\mu}}_{i}}\sim\text{Bern(\ensuremath{\Phi}(\ensuremath{\mathbf{\boldsymbol{\mu}}_{i}})})\,,
\end{equation}
where $\mathbf{y}_{i}\equiv(y_{i,1},y_{i,2}...,y_{i,6})^{'}$ is the
presence or absence of the six species at the $i^{\text{th}}$ site
($i=1,2,...,225)$. The parameter vector $\ensuremath{\mathbf{\boldsymbol{\mu}}_{i}}\equiv(\mu_{i,1},\mu_{i,2},...,\mu_{i,6})^{'}$
is mapped to a probability by applying the inverse probit link function,
$\Phi(\cdot)$, to each element. The parameter vector $\ensuremath{\mathbf{\boldsymbol{\mu}}_{i}}$
is specified using Eq. 4, but is simplified due to the single predictor
variable in this example
\begin{equation}
\mathbf{\boldsymbol{\mu}}_{i}=\mathbf{\boldsymbol{\alpha}}+x_{i}\mathbf{Z}\boldsymbol{\gamma}\,,
\end{equation}
where $\boldsymbol{\alpha}\equiv(\alpha_{1},\alpha_{2},...,\alpha_{6})^{'}$
is a vector that contains intercept parameters for each species and
$x_{i}$ is the recorded water depth at the $i^{\text{th}}$ site.
As in Eq. 4, the matrix $\mathbf{Z}$ has dimensions $J\times G$
where the $j^{\text{th}}$ row contains an indicator variable that
links the predictor variables to the guild-specific regression coefficients
contained within the vector $\boldsymbol{\gamma}\equiv(\gamma_{1,1},\gamma_{2,1},...,\gamma_{g,1})^{'}$.
For the intercept, we use the prior $\alpha_{j}\sim\text{N(0,1})$
because this results in a uniform distribution when $\alpha_{j}$
is transformed using the function $\Phi(\ensuremath{\alpha_{j}})$.
That is, using the prior $\alpha_{j}\sim\text{N(0,1})$ results in
$\Phi(\ensuremath{\alpha_{j}})\sim\text{unif(0,1})$.

We used a computationally efficient MCMC algorithm to fit the MSDM
to the aquatic vegetation data. This MCMC algorithm relies on an auxiliary
variable formulation of the binary regression model and has been used
to fit MSDMs to presence-absence data (\citealt{albert1993bayesian};
\citealt{wilkinson2019comparison}; \citealt{BB2L}, pgs. 246\textendash 253).
For each model fit, we use one chain and obtain 100,000 draws from
the posterior distribution, but retain every $10^{\text{th}}$ sample
to decrease storage requirements. Each model fit takes approximately
30 minutes using a desktop computer (see Appendix S1). We inspect
the trace plot for each parameter to ensure that the Markov chain
is sampling from the stationary distribution (i.e., the posterior
distribution). We discard the first 500 samples and use the remaining
9,500 samples for inference. 

The model-based recursive partitioning approach available in the R
function \texttt{lmtree(...)} and used to implement the TSP requires
the specification of a tuning parameter (hereafter \texttt{alpha};
see Appendix S1). This tuning parameter, \texttt{alpha}, must be between
zero and one and controls the splitting of parent nodes. Similar to
the tree generating stochastic process (see Section TREE SHRINKAGE
PRIOR), when \texttt{alpha$\,=1$} the tree has terminal nodes (guilds)
that contain only a single species. When \texttt{alpha$\,=0$} the
tree does not split and all six species are in the same guild. One
way to estimate the tree tuning parameter, \texttt{alpha}, is to find
the value that results in the most accurate predictions (\citealt{hobbs2015bayesian}).
For this example, we illustrate two approaches by estimating the predictive
accuracy of our models with scores that use either in-sample or out-of-sample
data. For the out-of-sample score, we use data from the 225 sites
that were not used to fit the model to calculate the log posterior
predictive density (LPPD). For the in-sample score, we calculate the
Watanabe-Akaike information criteria (WAIC). For comparison between
scoring approaches, we report $-2\times\text{LPPD}$ as we would expect
this and WAIC to result in similar numerical values (\citealt{gelman2014understanding}). 

For Bayesian models, computing in-sample scores like WAIC require
a measurement of a model's complexity. In non-hierarchical non-Bayesian
contexts, a model's complexity is the number of parameters in the
model. For example, using the MSDM applied to the aquatic vegetation
data, the number of parameters ranges from 7 (i.e., six intercept
parameters and one regression coefficient) to 12 (i.e., six intercept
parameters and six regression coefficients). Measuring the complexity
of our Bayesian MSDM with the TSP is more complicated than simply
counting the number of parameters because of the shrinkage effect
and because it is not obvious how many parameters are required to
estimate the number and species composition of the guilds. Fortunately,
when calculating Bayesian information criterion like WAIC, measure
of model complexity such as the ``effective number of parameters''
are available without any additional effort. To understanding how
the TSP controls the complexity of the MSDM, we report the effective
number of parameters using Eq. 12 from \citet{gelman2014understanding}
as \texttt{alpha} varies from zero to one.\vspace{0.66cm}

\begin{spacing}{1.9}

\subsection*{EXAMPLE 2: FISH ABUNDANCE}
\end{spacing}

Our objective for the second example is to demonstrate a different
and complex data type that necessitates a more sophisticated use of
the TSP and deeper ecological interpretation, similar to what ecologists
may encounter in practice. Here we use relative abundance data from
15 species of fish that were sampled at 83 sites, which included several
species that were present at only a small number of sites and are
considered rare or species of concern (Table 1). For example, at 79
of the 83 sites (95\%), a count of zero was recorded for river redhorse
(\textit{Moxostoma carinatum}; Table 1). At each site, numerous predictor
variables were measured, but for our analysis we use water temperature,
speed, and depth as these were most relevant to management of the
Upper Mississippi River System.

The fisheries data were collected over three time periods which included
period 1 (June 15 \textendash{} July 31), period 2 (August 1 \textendash{}
September 14), and period 3 (September 15 \textendash{} October 31).
The time periods are hypothesized to correspond to changes in habitat
use. For example, one hypothesis is that species composition of the
guilds and the regression coefficient estimates associated with water
temperature, speed, and depth may be different during each period.
In what follows, we demonstrate how to incorporate such dynamics into
MSDMs using the fisheries data.

For relative abundance data, \citet{johnson2017modeling} discuss
several commonly used MSDMs including a data model that has a marginal
distribution of
\begin{equation}
\mathbf{y}_{i}|\mathbf{z}_{i},\phi\sim\text{ZIP}(e^{\mathbf{z}_{i}},\phi)\,,
\end{equation}
where ``ZIP'' stands for zero-inflated Poisson, $e^{\mathbf{z}_{i}}$
is the expected value of the Poisson mixture component, and $\phi$
is the mixture probability (\citealt{BB2L}; pg. 320). We specify
the process model using Eq. 2 with $\boldsymbol{\Sigma\equiv}\sigma_{\varepsilon}^{2}\mathbf{I}$
and
\begin{equation}
\mathbf{\boldsymbol{\mu}}_{i}=\mathbf{\boldsymbol{\alpha}}+\left(\mathbf{x}_{i}^{'}\otimes\mathbf{Z}_{t}\right)\boldsymbol{\gamma}_{t},
\end{equation}
where the guild composition and regression coefficients vary over
the three sample periods as indicated by subscript $t$ such that
$\mathbf{Z}_{t}$ and $\boldsymbol{\gamma}_{t}$ corresponds to period
1 ($t=1$), period 2 ($t=2$), or period 3 ($t=3$). The vector $\mathbf{x}_{i}\equiv(x_{i,1},x_{i,2},x_{i,3})^{'}$
are the three predictor variables (water temperature, speed and depth)
measured at the $i^{\text{th}}$ site. In total, there are $135$
unique regression coefficients if the guilds contained only a single
species for all three periods, which would also be the case if the
standard MSDM (i.e., Eq. 3) with simple priors for the regression
coefficients was used. For the mixture probability, process model
variance, and intercepts we used the priors $\phi\sim\text{unif}(0,1)$,
$\sigma_{\varepsilon}^{2}\sim\text{IG}(2.01,1)$, and $\alpha_{j}\sim\text{N(0,10}^{3})$
respectively. 

For each model fit, we use one chain and obtain 30,000 draws from
the posterior distribution, but retain every $10^{\text{th}}$ sample
to decrease storage requirements. We inspect the trace plot for each
parameter to ensure the Markov chain is sampling from the stationary
distribution. We discard the first 500 samples and use the remaining
2,500 samples for inference.

The computational cost to implement the TSP will increase as the number
of species increase. In addition, allowing the number and species
composition of the guilds to vary over the three periods increases
the computational burden because the MSDM requires three TSPs instead
of one. As a results, we do not select the value of the tree tuning
parameter parameters that produces the best predictive model because
fitting the MSDM once to the fisheries data takes about 23 hours using
a desktop computer (see Appendix S2). Instead we choose a value of
the tree tuning parameters that favors a small number of guilds, which
we anticipate will make our results easier to share with policymakers.
Similar to the aquatic vegetation example, the tree tuning parameters
could be chosen to optimize the predictive ability of the model but
this would make the annotated computer code provided in Appendix S2
difficult for many readers to use without access to high-performance
computing resources.\vspace{0.66cm}

\section*{Results}
\begin{spacing}{1.9}

\subsection*{EXAMPLE 1: PRESENCE-ABSENCE OF AQUATIC VEGETATION}
\end{spacing}

The value of the tree tuning parameter that produces the most accurate
predictions was 0.6 when scored using LPPD and 0.025 when scored using
WAIC. In what follows, we present results obtained from the MSDM with
the TSP that used a value of the tree tuning parameter of 0.025 because
this value was optimal based on WAIC, results in only a slight reduction
in predictive accuracy when scored using LPPD, and yields a relatively
large reduction in the number of parameters which simplifies statistical
inference (Fig. 2). For the six species of aquatic vegetation, the
posterior distributions for the species-level regression coefficient,
obtained from Eq. 6, shows that the presence of wild celery has no
statistical relationship with water depth while the remaining five
species of vegetation exhibit a strong negative response to increases
in water depth (Fig. 3a,b). In addition to the posterior distribution
for the regression coefficients, derived quantities from the posterior
distribution of the tree can provide insightful inference. For example,
Fig 3c shows the posterior distribution of the number of terminal
nodes (i.e., guilds) and Fig 3d shows the expected value of the posterior
distribution of species co-occurrence within the same guild, both
of which were derived from the posterior distribution of the trees.
As another example, summaries of the posterior distribution of the
tree may be of interest. For example, Fig. 3e shows the most likely
tree, which is obtained from the mode of the posterior distribution
of the trees and can be used to infer which guild structure is most
likely. For the most likely tree, figure 3f shows the posterior distribution
of the two guild-level regression coefficients. Again, caution must
be taken when making inference from the guild-level regression coefficients
because these are conditional on a single tree and do not account
for guild uncertainty. For the aquatic vegetation data, however, the
inference does not differ among the species-level and guild-level
regression coefficients because the uncertainty in the number and
species composition of the guilds is relatively low.\vspace{0.66cm}

\begin{spacing}{1.9}

\subsection*{EXAMPLE 2: FISH ABUNDANCE}
\end{spacing}

\noindent Similar to the aquatic vegetation example, we show the posterior
distribution of the number of guilds (Fig. 4a,b,c), the expected value
of species co-occurrence within the same guild (Fig. 4d,e,f), and
the posterior distribution of the regression coefficients for each
species (Figs. 5 and S1). During period 1, our results indicate that
there are most likely two guilds (Fig 4a). The posterior distributions
of the regression coefficients show that eight species (bluegill,
bullhead minnow, emerald shiner, largemouth bass, mud darter, pugnose
minnow, tadpole madtom, and weed shiner; hereafter ``guild 1 during/in
period 1'') have large decreases in relative abundance when water
temperature and depth increases (Figs. 5 and S1). These results suggest
guild 1 during period 1 is more abundant in shallower side channels
and off-channel habitats compared to the main channel, where depth
is maintained at a minimum of 2.75 m for navigation. Of the species
in guild 1 during period 1 many are nest spawners (e.g., bluegill,
bullhead minnow, largemouth bass, pugnose minnow, tadpole madtom)
and our results correspond with habitat requirements for successful
reproduction. During period 1, the remaining seven species (golden
redhorse, longnose gar, river redhorse, river shiner, shorthead redhorse,
smallmouth bass, spotfin shiner; hereafter ``guild 2 during/in period
1'') have posterior distributions of the regression coefficients
that indicate slight decrease in relative abundance when water temperature
or depth increases, however, here is a non-negligible probability
that these responses could be close to zero or even positive (Figs.
5 and S1). This suggest that during period 1 species in guild 2 are
found over a broader range of conditions when compared to species
in guild 1. Many of these species are classified as fluvial dependent
or fluvial specialists, which are either generally found in lotic
environments or require flowing water for some part of their life
history (\citealt{galat2001conserving,lifehistory}). 

During period 2, our results indicate that there is most likely a
single guild that contains all 15 species of fish (Fig 4b). The posterior
distributions of regression coefficients for all species generally
have an expected value near zero, suggesting random associations with
water depth, speed, and temperature (Fig. 5 and S1). The lack of a
response to water depth, speed, and temperature during period 2 may
be a result of active foraging in which species are moving in search
of food. 

During period 3, our results indicate that there are most likely two
guilds (Fig 4c). The posterior distributions of the regression coefficients
associated with depth indicate a negative response for five species
(bluegill, largemouth bass, pugnose minnow, tadpole madtom, weed shiner;
hereafter ``guild 1 during/in period 3''), which suggests an avoidance
of the main channel (Figs. 5 and S1). In addition, our results show
a negative response to water speed for the species in guild 1 during
period 3 which suggest associations with lentic habitats such as backwater
lakes. As temperatures decrease during period 3 (September 15 \textendash{}
October 31), many of the species in guild 1 are known to move into
slow-moving backwaters to minimize energy expenditures during winter
conditions. During period 3, six species (bullhead minnow, emerald
shiner, river redhorse, shorthead redhorse, smallmouth bass, spotfin
shiner; ``guild 2 during/in period 3'') had a slightly positive
association with water speed and negative association with water temperature
(Figs. 5 and S1). These results suggest that species in guild 2 during
period 3 are more likely to inhabit the main channel and side channels
during this period than off-channel habitats. During period 3, there
are four species (golden redhorse, longnose gar, mud darter, and river
shiner) whose posterior distributions of regression coefficients do
not clearly delineate guild association. These results suggest either
a high level of guild uncertainty, insufficient data, or weaker associations
with the variables examined.

In summary, the number and composition of guilds changed over the
three periods, suggesting shifts in habitat associations and species
interactions through time. Whether these shifts were related to specific
requirements for spawning, foraging, or overwintering would require
additional field research designed to answer such questions, but our
results are consistent with our understanding of fluvial dependence
and general habitat guilds. 

\section*{Discussion}

A benefit of hierarchical Bayesian modeling is that statistical models
can be easily customized to match the goals of a study. In what follows,
we discuss modifications to the data and/or process models that have
been commonly used to accommodate different types and quality of data.
The modifications we discuss can be incorporated without making major
changes to the basic framework we presented. We begin by discussing
modifications to the data model. For our aquatic vegetation and fisheries
data examples we used data models that were appropriate for presence-absence
and relative abundance data, respectively. Other types of ecological
data may require the specification of different distributions for
the data. For example, a beta distribution is often used for plant
cover data, which are usually reported as percentages (\citealt{wright2017statistical};
\citeauthor{damgaard2019using} \textit{in press}). Our approach,
using the process model in Eq. 3 and TSP, could be used to develop
a MSDM for plant cover data by specifying a beta distribution as the
data model. Similarly, selecting a data model that matches key characteristics
of the observed data (e.g., the support) is a common model building
practice and applications using the TSP are straightforward (\citealt{hobbs2015bayesian,BB2L}). 

Many types of ecological data are measured with error. For example,
presence-absence data may contain false-negatives (\citealt{tyre2003improving,martin2005zero,gray2013influences}).
For MSDMs, accounting for measurement error in the data is an important
component of the model building process (\citealt{beissinger2016incorporating,warton2016extending,tobler2019joint}).
For example, multi-species occupancy models have been developed to
account for false-negatives in presence-absence community data (\citealt{dorazio2005estimating}).
Model selection is challenging for multi-species occupancy models
(\citealt{broms2016model}), however, using the approaches demonstrated
in this paper, it is straightforward to implement the TSP for the
multi-species occupancy model. This would involve adding a fourth
level to the hierarchical MSDM in Eqs. 2\textendash 3 that accounts
for the possibility of false-negatives in the presence-absence data
(\citealt{BB2L}, ch. 23). Using the information provided in Appendix
S1 in concert with \citet{dorazio2012gibbs} or \citeauthor{BB2L}
(2019; ch. 23) would yield an efficient implementation that requires
a minimal amount of additional programming.

As noted in the introduction, a main focus of methodological development
within the literature on MSDMs has been on parametrizing the variance-covariance
matrix, $\boldsymbol{\Sigma}$, for the process model in Eq. 2. In
previous studies the variance-covariance matrix was specified to account
for residual autocorrelation due to space, time, or biotic interactions.
These modifications to the variance-covariance matrix can be incorporated
into a MSDM that uses the TSP. For example, latent variables parameterizations
have been used to induce dependence among species and estimate $\boldsymbol{\Sigma}$
(\citealt{walker2011random,warton2015so,hui2016boral}). As another
example, \citet{johnson2017modeling} note that spatial or temporal
autocorrelation can be accounted for by including basis functions
to estimate $\boldsymbol{\Sigma}$ (\citealt{hefley2017basis}). In
both the latent variables parameterization and basis function approach,
the matrix $\boldsymbol{\Sigma}$ is implicitly specified using a
``random effects'' or ``first-order representation'' (\citealt{hefley2017basis};
\citealt{wikle2019spatio}, pgs. 156\textendash 157).

The TSP is easy to program using the approximate Gibbs sampling approach
from \citet{shaby2012embedding} and requires only a few additional
lines of code when compared to Bayesian MSDMs that use simple priors.
The computational burden to implement the TSP, however, is considerably
larger which may increase the run-time required to fit MSDMs to data.
For example, fitting independent models for single-species or MSDM
with simple priors to data is faster because the number and species
composition of the guilds does not have to be estimate and it may
be easy to parallelize the computational procedure (\citealt{BB2L},
ch. 19; \citeauthor{hooten2018making} \textit{in press}). In general,
the TSP will become more computationally challenging to implement
as the number of species increases. For example, our implementation
which uses a model-based recursive partitioning approach available
in the R function \texttt{lmtree(...)} has a computational complexity
that increase nonlinearly with the number of species. As a result
fitting a MSDM to data for a small number of species is relatively
fast as demonstrated by our example with six species of aquatic vegetation,
however, model fitting requires much more time for a larger number
of species as demonstrated by our example with 15 species of fish.
The increase in computational time is caused by the increase in the
number of possible guilds. Based on our experience and using the computational
approaches presented in this paper, the TSP becomes infeasible to
fit to data that contains more than roughly twenty species, however,
feasibility will vary by data set and computer resources. Although
this is a current limitation of the TSP, other Bayesian MSDMs that
incorporate the guild concept face similar challenges (e.g., \citealt{johnson2017modeling})
which demonstrates the need for more efficient algorithms for ``big''
ecological communities. 

Currently we suggest two approaches to alleviate the computational
burden when applying our approach to a large number of species. The
first suggestion is to choose off-the-shelf tree methods that use
multiple processors (or cores). For example, we used the R function
\texttt{lmtree(...)}which allows users to choose the number of processor
cores for tree related computations. In practice, implementation that
exploits multiple processors should reduce the time required to fit
a MSDM with the TSP to data, however, gains will vary by application.
The second approach to reduce the computational burden involves using
values of the tree hyperparameters that favor a small number of guilds.
In studies with a large number of species, choosing a value of the
tree hyperparameters that results in a small number of guilds may
reduce the predictive accuracy of the model, however, model fitting
will likely be quicker. In addition a small number of guilds may be
desirable for studies that contain a large number of species because
interpretation of the results for a larger number of guilds may be
challenging. Both techniques, using multiple cores and reducing the
number of guilds, were illustrated in our fisheries data example (see
Appendix S2). As with most computationally demanding statistical methods,
advances may soon obviate the current challenges.

Model development, implementation, and selection for data from ecological
communities is difficult because a high level of complexity is desired
which can be achieved by including numerous parameters. For example,
commonly used MSDMs are specified so that each species has a unique
regression coefficient for each predictor variable (e.g., \citealt{wilkinson2019comparison}).
As a result it is common practice to overparameterize MSDMs, which
can degrade predictive accuracy and produce results that are difficult
to interpret and communicate. We illustrated how to reduce the number
of parameters in commonly used Bayesian MSDMs by replacing simple
prior distributions for regression coefficients with a TSP. The TSP
reduces model complexity while incorporating ecological theory, expanding
inference, and can increase the predictive accuracy and interpretability
of Bayesian MSDMs. 

\vspace{0.66cm}

\section*{Acknowledgements}

The U.S. Army Corps of Engineers' Upper Mississippi River Restoration
Program LTRM element is implemented by the U.S. Geological Survey,
Upper Midwest Environment Sciences Center, in cooperation with the
five Upper Mississippi River System states of Illinois, Iowa, Minnesota,
Missouri, and Wisconsin. The U.S. Army Corps of Engineers provides
guidance and has overall Program responsibility. The authors acknowledge
support for this research from USGS G17AC00289. Use of trade, product,
or firm names does not imply endorsement by the U.S. Government.\vspace{0.66cm}

\section*{Data accessibility}

The aquatic vegetation and fisheries data are publicly available from
\citet{UMRR}, but the specific subsets used in this study will be
archived in the Dryad Digital Repository. For the purpose of peer
review, the data have been submitted as supporting material as a compressed
file labeled Data.zip.

\renewcommand{\refname}{\vspace{0.1cm}\hspace{-0cm}\selectfont\large References}\setlength{\bibsep}{0pt}

\bibliographystyle{apa}
\bibliography{refs}
\vspace{-0.5cm}

\section*{Supporting Information}

\begin{singlespace}
\noindent Additional Supporting Information may be found in the online
version of this article.
\end{singlespace}

\noindent \textbf{Appendix S1}

\noindent Tutorial with R code to reproduce the aquatic vegetation
example and Figs. 2 and 3.

\noindent \textbf{Appendix S2 }

\noindent Tutorial and R code to reproduce the fisheries example and
Figs. 4, 5, and S1.

\noindent \textbf{Appendix S3}

\noindent Supporting Fig. S1.\pagebreak{}

\noindent \textbf{Fig. 1.} An example of a binary tree that is used
to partition an ecological community that contains six species of
aquatic vegetation into three guilds. The tree shrinkage prior uses
a binary tree to determine which species share the same value of the
regression coefficients, $\gamma_{g,k}$, for the predictor variable
water depth. In this example, the standard priors for multi-species
distribution model forces each species to have a unique regression
coefficient. The binary tree identifies the species composition and
number of guilds. For the species within a guild, a single regression
coefficient is used to model the predictor variable water depth, which
reduces the number of regression coefficients from six to three.\bigskip{}

\noindent \textbf{Fig. 2.} Results from fitting multi-species distribution
models to presence-absence data for six species of aquatic vegetation
using the predictor variable water depth. Panel a shows how the predictive
score, $-2\times\text{LPPD}$ (where LPPD is the log posterior predictive
density calculated from out-of-sample data), changes as the tree tuning
parameter varies from zero to one. Panel b shows WAIC, which is similar
to $-2\times\text{LPPD}$, but uses in-sample data. For both WAIC
and $-2\times\text{LPPD}$, a lower score indicates more accurate
predictions, which can be optimized by varying the tree tuning parameter
from zero to one. Panel c shows the posterior mode of the number of
guilds as the tree tuning parameter varies from zero to one while
panel d shows the effective number of parameters. All models contain
six species-specific intercept parameters, but the number of regression
coefficients varies from one to six depending on the value of the
tree tuning parameter. When the tree tuning parameter is set to one
each species is assigned to a different guild, resulting in the standard
species-specific MSDM (Eq. 3) with six regression coefficients for
a total of 12 parameters. When the tree tuning parameter is set to
zero the result is a single guild that contains all six species and
one regression coefficient for a total of seven parameters. Panel
e shows the expected value of the species-level regression coefficients
associated with depth as the tree tuning parameter varies from zero
to one. As the tree tuning parameter decreases to zero, the species-level
regression coefficient estimates are shrunk towards $\approx-0.5$,
which is the regression coefficient estimate when all species are
in the same guild.\bigskip{}

\noindent \textbf{Fig. 3.} Posterior distributions and summaries obtained
from fitting the multi-species distribution model to presence-absence
data from 225 sites with a tree tuning parameter value of $0.025$.
Panel a shows the posterior distributions of the regression coefficient
for all six species associated with the predictor variable water depth.
Panel b shows the expected value of the probability of species presence
as water depth varies from zero to three meters. Panel c shows the
posterior distribution for the number of guilds. Panel d shows the
expected value of the posterior distribution of species co-occurrence
within the same guild. Panel e shows the mode of the posterior of
the tree structure (i.e., the most probable guild structure). Panel
f shows the posterior distributions of the regression coefficient
for the two guilds from panel e. The color-coded tick marks on the
horizontal axis of panels a and f are the expected value of the corresponding
posterior distribution. The square-bracket notation is used to represent
posterior density functions. \bigskip{}

\noindent \textbf{Fig. 4.} Results obtained from fitting a multi-species
distribution model to abundance data from 83 sites and 15 fish species.
We specified the model to allow the number and species compositions
of guilds to vary over three time periods. Panels a, b, and c show
the posterior distribution for the number of guilds in time period
1 (June 15-July 31), period 2 (August 1- September 15), and period
3 (September 16 - October 31) respectively. Panels d, e, and f show
the expected value of the posterior distribution of species co-occurrence
within the same guild during periods 1, 2, and 3 respectively.\bigskip{}

\noindent \textbf{Fig. 5.} Posterior distributions of regression coefficients
obtained from fitting our multi-species distribution model to relative
abundance data from 83 sites that included 15 species of fish (see
Table 1 for species list). Panels a, b, and c show the posterior distributions
of regression coefficients for the variables water temperature, speed,
and depth for bluegill respectively, which was the most common species
in our study. Panels d, e, and f show the posterior distributions
of regression coefficients for the variables water temperature, speed,
and depth for river redhorse respectively, which was the least common
species in our study. The posterior distribution for river redhorse
are bimodal due to uncertainty in the guild membership of the species.
The color-coded tick marks on the horizontal axis shows the expected
value of the corresponding posterior distribution. The square-bracket
notation is used to represent posterior density functions. The predictor
variables variables water temperature, speed, and depth were centered
and standardized prior to fitting the model and had a standard deviation
of 3.8, 0.16, and 0.44 respectively. Fig. S1 in Appendix S2 contains
posterior distributions for all 15 species. 

\pagebreak{}
\begin{figure}[H]
\begin{centering}
\includegraphics[scale=0.5]{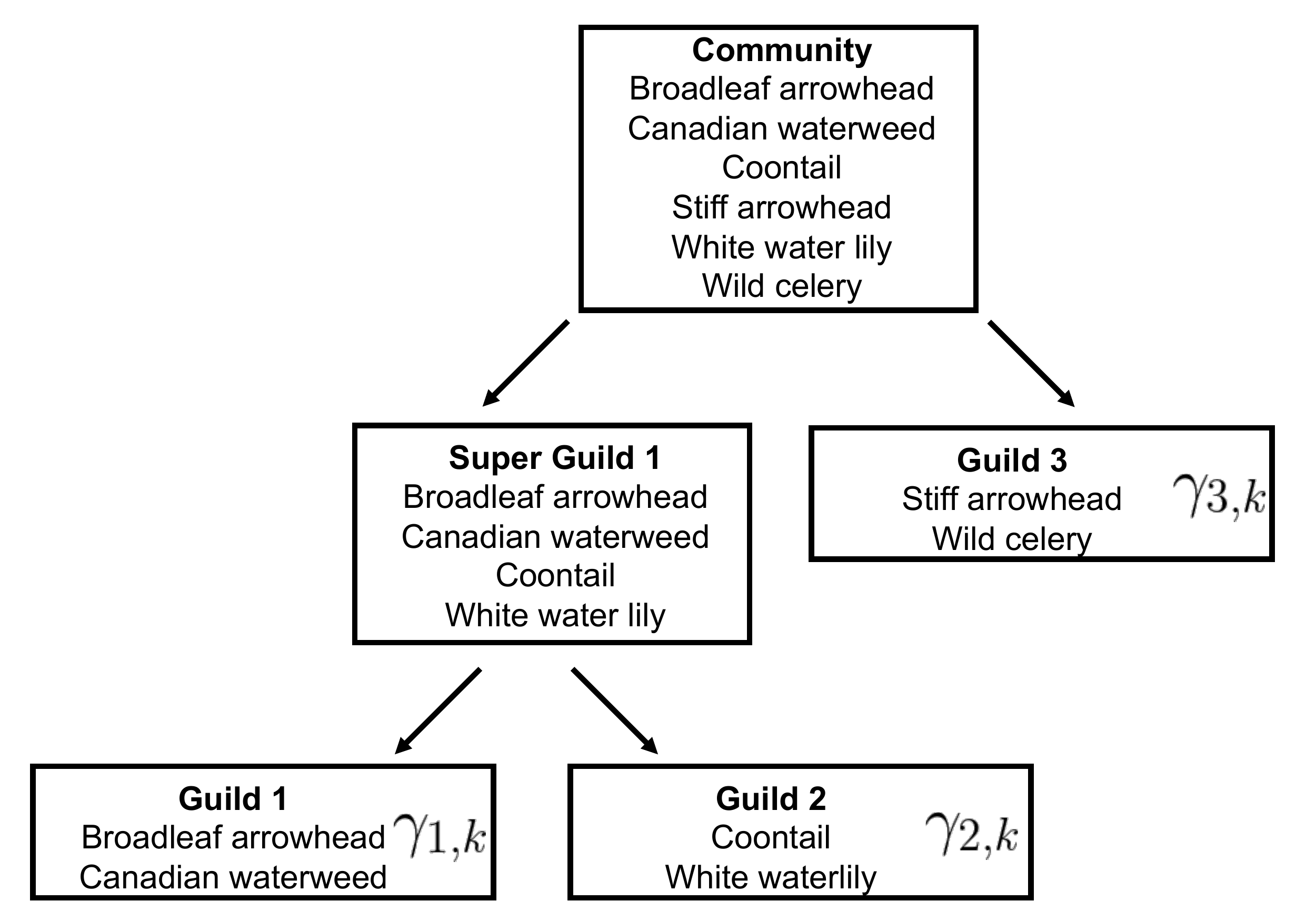}
\par\end{centering}
\caption{}
\end{figure}
\begin{figure}[H]
\begin{centering}
\includegraphics[scale=1.15]{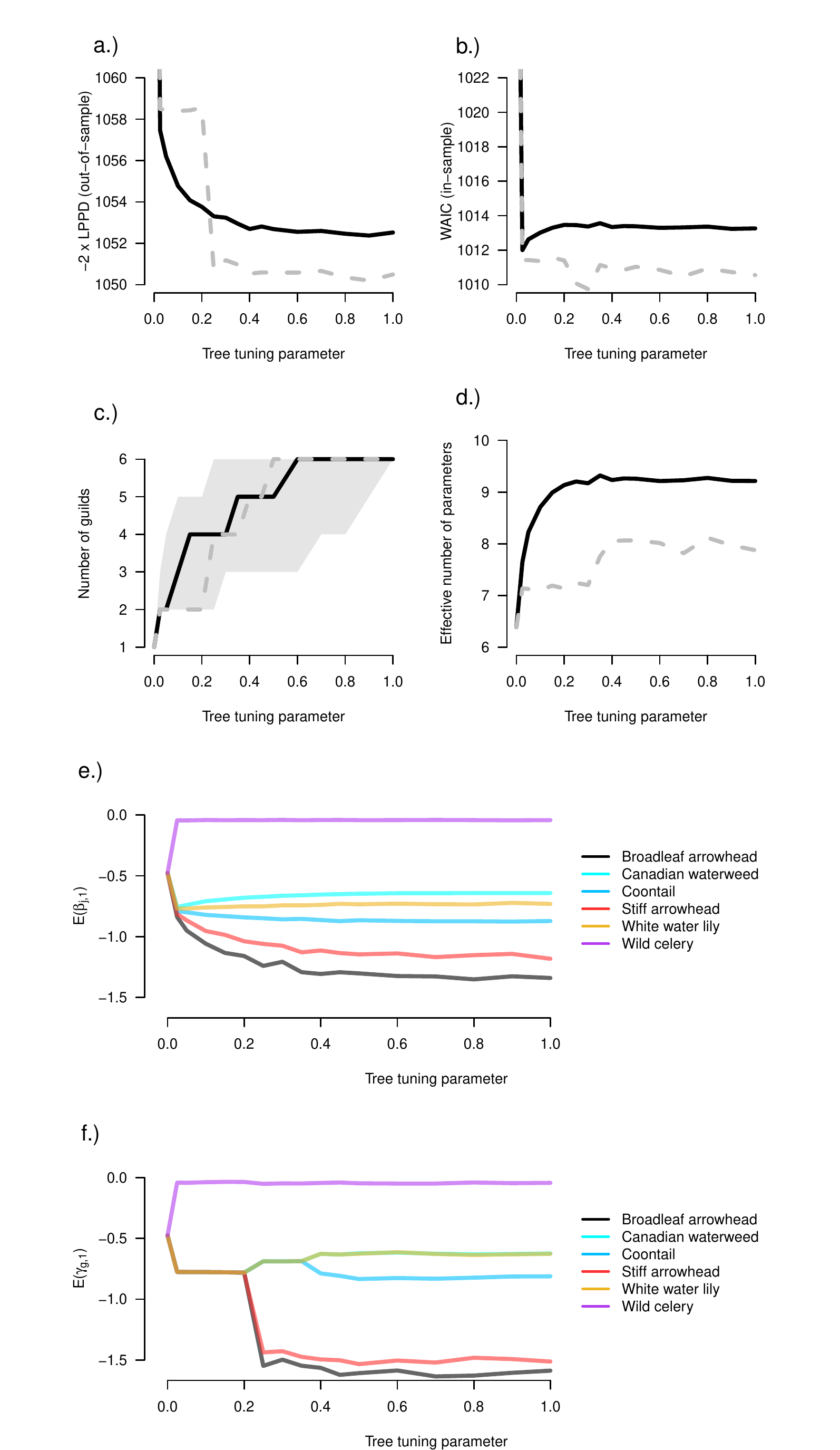}
\par\end{centering}
\caption{}
\end{figure}
\begin{figure}[H]
\begin{centering}
\includegraphics[scale=0.8]{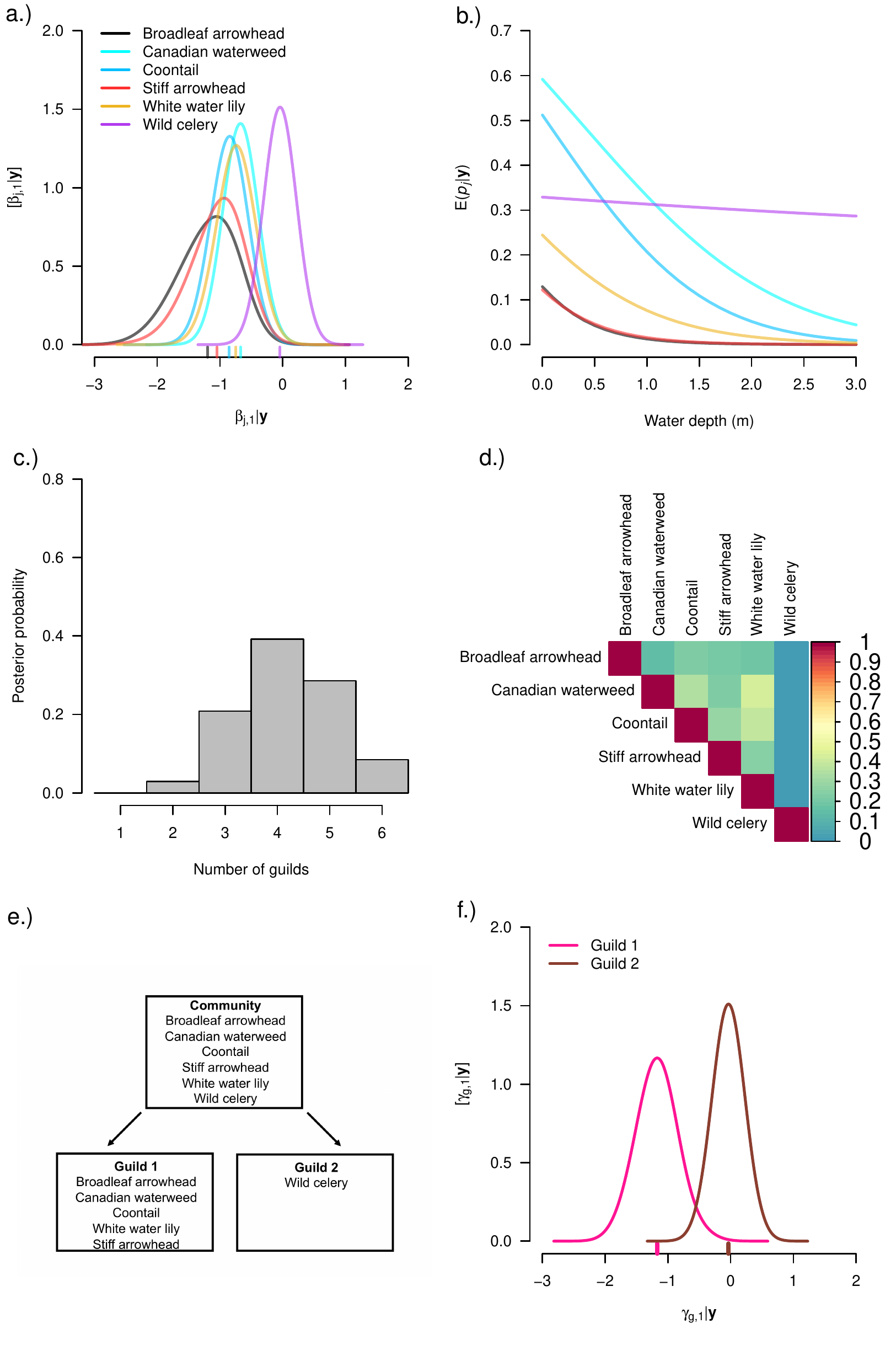}
\par\end{centering}
\caption{}
\end{figure}
\begin{landscape}
\begin{figure}[H]
\begin{centering}
\includegraphics[scale=0.9]{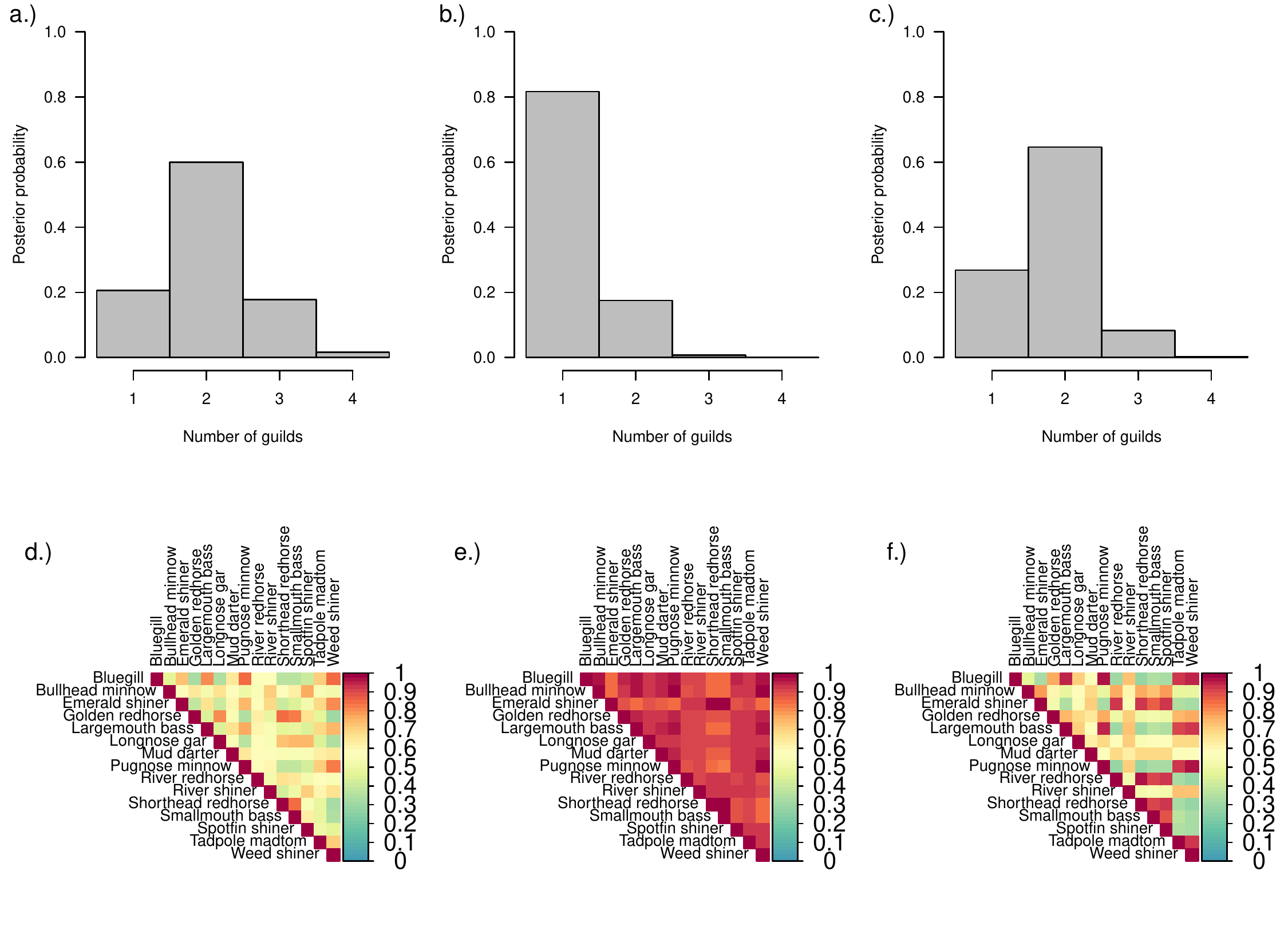}
\par\end{centering}
\caption{}
\end{figure}
\begin{figure}[H]
\begin{centering}
\includegraphics{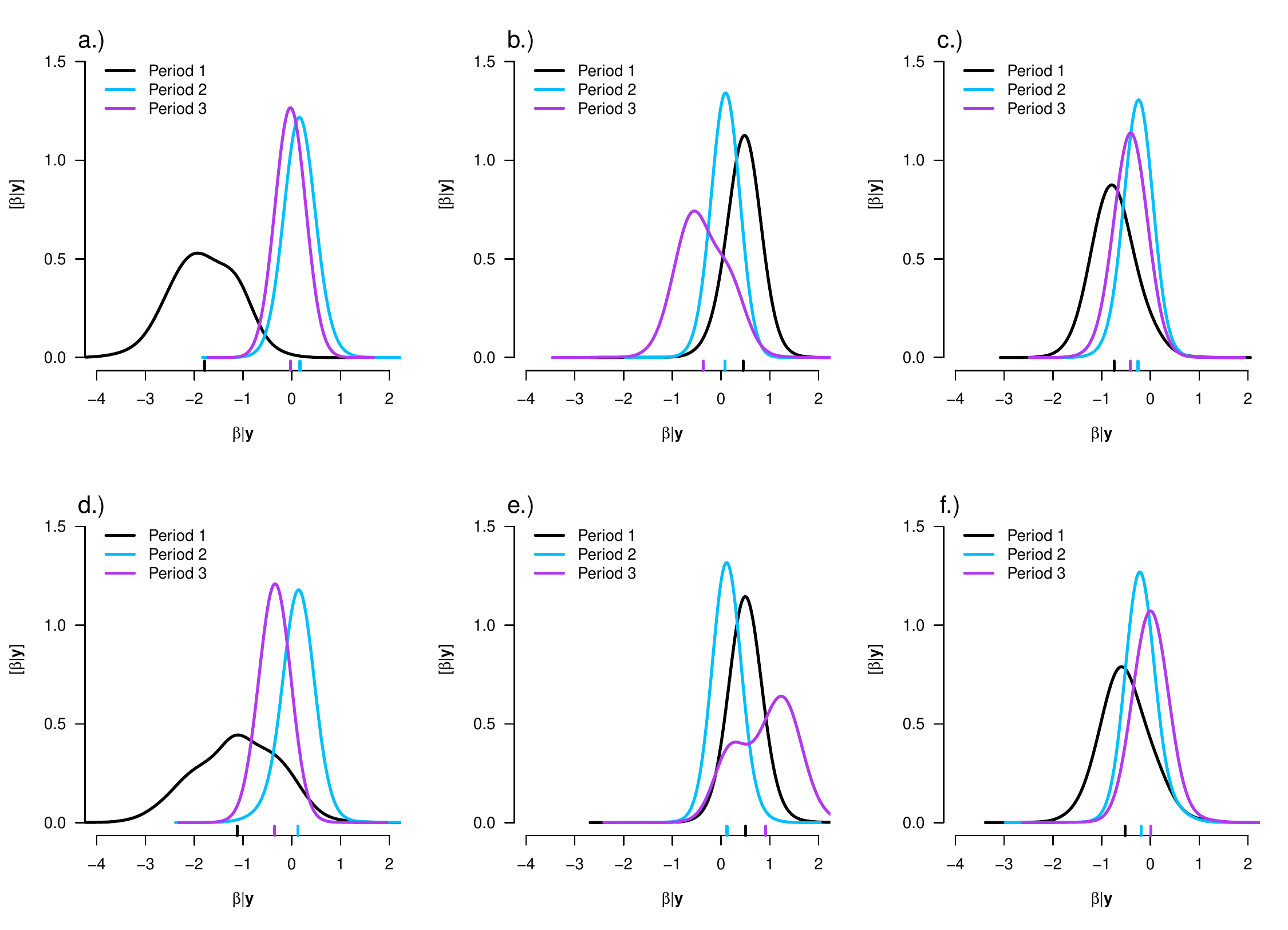}
\par\end{centering}
\caption{}
\end{figure}

\pagebreak{}

\end{landscape}

\begin{flushleft}
\textbf{Table 1}. Common name, species name, and percent of 83 sites
with counts greater than zero for 15 species of fish used in our example. 
\par\end{flushleft}

\begin{center}
\begin{tabular}{ccc}
\hline 
Common name & Species & Percent of sites with counts greater than zero\tabularnewline
\hline 
Bluegill & \textit{Lepomis macrochirus} & 81\%\tabularnewline
Bullhead minnow & \textit{Pimephales vigilax} & 45\%\tabularnewline
Emerald shiner & \textit{Notropis atherinoides} & 48\%\tabularnewline
Golden redhorse & \textit{Moxostoma erythrurum} & 53\%\tabularnewline
Largemouth bass  & \textit{Micropterus salmoides} & 84\%\tabularnewline
Longnose gar  & \textit{Lepisosteus osseus} & 22\%\tabularnewline
Mud darter & \textit{Etheostoma asprigene} & 6\%\tabularnewline
Pugnose minnow & \textit{Opsopoeodus emiliae} & 27\%\tabularnewline
River redhorse & \textit{Moxostoma carinatum} & 5\%\tabularnewline
River shiner & \textit{Notropis blennius} & 17\%\tabularnewline
Spotfin shiner & \textit{Cyprinella spiloptera} & 71\%\tabularnewline
Shorthead redhorse & \textit{Moxostoma macrolepidotum} & 62\%\tabularnewline
Smallmouth bass & \textit{Micropterus dolomieu} & 49\%\tabularnewline
Tadpole madtom & \textit{Noturus gyrinus} & 8\%\tabularnewline
Weed shiner & \textit{Notropis texanus} & 54\%\tabularnewline
\hline 
\end{tabular}
\par\end{center}


\end{flushleft}
\end{document}